\documentclass[a4paper]{article}
\usepackage{bbding}
\usepackage[T1]{fontenc}
\usepackage[utf8]{inputenc}
\usepackage{aurical}
\usepackage{huncial}
\usepackage{linearb}
\usepackage{textcomp}
\usepackage{amsmath}
\usepackage{amssymb}
\usepackage{amsfonts}
\usepackage{amssymb,epic,eepic}
\usepackage[amsmath,amsthm,thmmarks]{ntheorem}
\usepackage[usenames,dvipsnames]{pstricks}
\usepackage{epsfig}
\usepackage{pst-grad}
 \usepackage{pst-plot}

\newtheorem{theorem}{Theorem}[section]
\newtheorem{lemma}[theorem]{Lemma}

\begin{document}
\title{Classical-Quantum Arbitrarily Varying Wiretap Channel}
\author{Vladimir Blinovsky, Minglai Cai}
\maketitle
\footnote{Project "Informationstheorie des Quanten-Repeaters" supported by the
Federal Ministry of Education and Research (Ref. No. 01BQ1052)}
\begin{abstract}
We derive a lower bound on the secrecy  capacity of classical-quantum
arbitrarily varying wiretap channel for both the case with and
without channel state information at the transmitter.
\end{abstract}
\section{Introduction}The arbitrarily varying channel models transmission over a channel
with an state that can change over time. We may
interpret it as a channel with an evil jammer.  The arbitrarily
varying channel was first introduced by Blackwell, Breiman, and
Thomasian in \cite{Bl/Br/Th}. The wiretap channel models communication
with security. It was first introduced by Wyner in \cite{Wyn}. We
may interpret it as a channel with an evil eavesdropper. The
arbitrarily varying wiretap channel models transmission with both a
jammer and an eavesdropper. Its capacity has been
determined by Bjelakovi\'{c}, Boche, and Sommerfeld in \cite{Bj/Bo/So}.

A quantum  channel is a channel which can transmit both classical
and quantum information. In this paper, we consider the capacity of
quantum channels to carry classical information, or equivalently,
the capacity of a classical quantum channels. The classical capacity
of quantum channels has been determined by Holevo in \cite{Ho}.
 A
classical-quantum  channel with a jammer is called a
classical-quantum  arbitrarily varying  channel, its capacity has been
determined by Ahlswede and Blinovsky in \cite{Ahl/Bli}.
 Bjelakovi\'{c},  Boche,  Jan\ss en, and  N\"otzel gave an alternative proof
and a proof of the
strong converse in \cite{Bj/Bo/Ja/No}. A
classical-quantum  channel with an eavesdropper is called a
classical-quantum wiretap  channel, its capacity has been determined by
Devetak in \cite{Dev}, and by  N. Cai,  Winter, and  Yeung in
\cite{Ca/Wi/Ye}.

 A
classical-quantum  channel with both a jammer and an eavesdropper is
called a classical-quantum  wiretap  channel, it is defined as a
pair of double indexed finite set of density operators
$\{(\rho_{x,t}, \sigma_{x,t}) :x\in \mathcal{X}, t \in \Theta\}$
with common input alphabet $\mathcal{X}$ connecting a sender with
two receivers, one legal and one wiretapper, where $t$ is called a
state of the channel pair. The legitimate receiver accesses the
output of the first channel $\rho_{x,t}$ in the pair $(\rho_{x,t},
\sigma_{x,t})$, and the wiretapper observes the output of the second
part $\sigma_{x,t}$ in the pair $(\rho_{x,t}, \sigma_{x,t})$,
respectively, when a state $t$, which varies from
symbol to symbol in an  arbitrary manner, governs both the legitimate receiver's
channel and the wiretap  channel. A code for the
channel conveys information to the legal receiver such that the
wiretapper knows nothing about the transmitted information. This is
a generalization of model of
classical-quantum compound wiretap channels in \cite{Ca/Ca/De} to
the case when the channel states are not stationary, but can change over the time.

We will be dealing with two communication scenarios. In the first
one only the transmitter is informed about the index $t$ (channel
state information, or simply CSI, at the transmitter), while in the second,
the legitimate users have no information about that index at all (no
CSI).

\section{Definitions}\label{def}
Let $\mathcal{X}$ be a finite set (the set of code symbols). Let
$\Theta$ := $\{1,\cdots,T\}$ be finite set (the set of channel
states). Denote  the set of the (classical)
messages by $\{1,\cdots,J_n\}$. Define the classical-quantum arbitrarily varying wiretap
channel  by a pair of double indexed finite set of density operators
$\{(\rho_{x,t}, \sigma_{x,t}) :x\in \mathcal{X}, t \in \Theta\}$ on
$\mathbb{C}^d$. Here the first family represents the communication
link to the legitimate receiver while the output of the latter is
under control of the wiretapper.

One important notation in \cite{Ahl/Bli} is the symmetrizable classical-quantum
arbitrarily varying  channel. We say
$\{\rho_{x,t}:x\in\mathcal{X},t\in\Theta\}$ is symmetrizable if
there exists a parameterized set of distributions
$\{U(t|x):x\in\mathcal{X}\}$ on $\Theta$ such that for all
$x,x'\in\mathcal{X}$ the following equalities are valid:
\[\sum_{t\in\Theta}U(t|x)\rho_{x',t}=\sum_{t\in\Theta}U(t|x')\rho_{x,t}\]

For any probability distribution $P\in\mathcal{P}$ and positive
$\delta$ denote $\mathcal{T}^n_{P,\delta}$ the $\delta$-typical set
in sense of \cite{Cs/Ko}.

For a state $\rho$, the von Neumann entropy  is defined as
\[S(\rho) := -\mathrm{tr} (\rho \log \rho)\text{ .}\]
Let $P$ be a probability distribution over a finite set $J$, and
$\Phi := \{\rho(x) : x\in J\}$  be a set of states labeled by
elements of $J$. Then the  Holevo $\chi$ quantity is
defined as
\[\chi(P,\Phi):= S\left(\sum_{x\in J} P(x)\rho(x)\right)-
\sum_{x\in J} P(x)S\left(\rho(x)\right)\text{ .}\]

 A (deterministic) quantum code $C$
of cardinality $J_n$ and length $n$ is a set of pairs
$\{(c_{j}^n,D_j) : j= 1,\cdots J_n\}$, where $c_{j}^n = (c_{j,1},
c_{j,2}, \cdots, c_{j,n}) \in \mathcal{X}^n$, and $\left\{D_j: j=
1,\cdots J_n\right\} $ is a collection of positive semi-definite
operators which is a resolution of the identity in
$(\mathbb{C}^d)^{\otimes n}$, i.e. $\sum_{j=1}^{J_n} D_j =
id_{(\mathbb{C}^d)^{\otimes n}}$.

A non-negative number $R$ is an achievable secrecy rate for the
classical-quantum arbitrarily varying wiretap channel if for every
$\epsilon>0$, $\delta>0$, $\zeta>0$ and sufficiently large $n$ there
exist a code $C = \{(c_{j}^n,D_j) : j= 1,\cdots J_n\}$  such that
\[\frac{\log J_n}{n} > R-\delta\text{ ,}\]
\[\max_{t^n \in \Theta^n} P_e(C, t^n) < \epsilon\text{ ,}\]
\[\max_{t^n\in\Theta^n}\frac{1}{n}
\chi\left(W,\{\sigma_{c_j^{n},t^n}:j=1,\cdots,J_n\}\right) <
\zeta\text{ ,}\] where $W$ is an uniformly distributed random
variable with values in $\{1,\cdots J_n\}$. Here $P_e(C, t^n)$ (the
average probability of the decoding error of a deterministic code
$C$, when the state (sequence of states) of the classical-quantum
arbitrarily varying wiretap channels is $t^n = (t_1, t_2, \cdots ,
t_n)$) is defined as follows
\[ P_e(C, t^n) := 1- \frac{1}{J_n} \sum_{j=1}^{J_n} \mathrm{tr}(\rho_{c_j^n,t^n} D_j)\text{ ,}\]
where $\rho_{c_j^n,t^n}:=\rho_{c_{j,1},t_1} \otimes
\rho_{c_{j,2},t_2} \otimes \cdots  \otimes \rho_{c_{j,n},t_n}$.

A non-negative number $R$ is an achievable secrecy rate for the
classical-quantum arbitrarily varying wiretap channel with channel
state information (CSI) at the transmitter if for every
$\epsilon>0$, $\delta>0$, $\zeta>0$ and sufficiently large $n$ there
exist for every $t^n$ a code $C^{t^n}=\{(c_{j,t^n}^{n},D_j) :  j= 1,\cdots
J_n\}$, where $c_{j,t^n}^n = (c_{j,1,t^n}, c_{j,2,t^n}, \cdots,
c_{j,n,t^n}) \in \mathcal{X}^n$, such that
\[\frac{\log J_n}{n} > R-\delta\text{ ,}\]
\[\max_{t^n \in \Theta^n} P_e^{CSI}(C^{t^n}, t^n) < \epsilon\text{ ,}\]
\[\max_{t^n\in\Theta^n}\frac{1}{n}\chi\left(W,
\{\sigma_{c_{j,t^n}^{n},t^n}:j=1,\cdots,J_n\}\right)
< \zeta\text{ .}\] Here $P_e^{CSI}(C^{t^n}, t^n)$
 is defined as follows:
\[ P_e^{CSI}(C^{t^n}, t^n) := 1- \frac{1}{J_n} \sum_{j=1}^{J_n} \mathrm{tr}(\rho_{c_{j,t^n}^{n},t^n} D_j)\text{ ,}\]
where  $\rho_{c_{j}^{t^n},t^n}:=\rho_{c_{j,t^n}^{n},t_1} \otimes
\rho_{c_{j,1,t^n},t_2} \otimes \cdots  \otimes
\rho_{c_{j,n,t^n},t_n}$.

One tool we will use is the random
quantum code, which we will define now. Let $\Lambda = \left(\mathcal{X}^n\times
\mathcal{B}((\mathbb{C}^d)^{\otimes n})\right)^{J_n}$. A random
quantum code $(\{C^{\gamma}:\gamma\in \Lambda\},G)$ consists of the
 family of sets of $J_n$ pairs $C^{\gamma}=\{(c_j^{n,\gamma},D_j^{\gamma}):
 j=1,\cdots,J_n\}_{\gamma\in \Lambda}$, where $c_j^{n,\gamma}=(c_{j,1}^{n,\gamma}\cdots
 c_{j,n}^{n,\gamma}) \in \mathcal{X}^n$ and $\sum_{j=1}^{J_n}D_j^{\gamma}=id_{(\mathbb{C}^d)^{\otimes n}}$,
 together with a distribution $G$ on $\Lambda$. The average
probability of the decoding error is defined as follows
\[P_{er} := \inf_{G}
\max_{t^n\in\Theta^n}\int_{\Lambda}P_{e}(C^{\gamma},t^n)d
G(\gamma)\text{ .}\]

A non-negative number $R$ is an achievable secrecy rate for the
classical-quantum arbitrarily varying wiretap channel under random
quantum coding if  for every
 $\delta>0$, $\zeta>0$, and  $\epsilon>0$, if $n$ is sufficiently large,
 we can find a $J_n$ such that
\[\frac{\log J_n}{n} > R-\delta\text{ ,}\]
\[ P_{er} < \epsilon\text{ ,}\]
\[\max_{t^n\in\Theta^n}\max_{\gamma\in\Lambda}\frac{1}{n}
\chi\left(W,\{\sigma_{c_j^{n,\gamma},t^n}:j=1,\cdots,J_n\}\right)
< \zeta\text{ .}\]

 Denote $P^n(x^n) := P(x_1) P(x_2)\cdots  P(x_n)$.
The following facts hold: (cf. \cite{Wil})\vspace{0.15cm}

Let $\mathcal{X}'$ be a finite set and for any $x\in\mathcal{X}'$,
 $\varsigma_x$ be a density operator  on $\mathbb{C}^d$.
For any distribution $P$ on $\mathcal{X}'$ and
$x^n \in\mathcal{T}^n_P $ let $\varsigma_{x^n} :=
\varsigma_{x_1}\otimes\varsigma_{x_2}\otimes\cdots\otimes\varsigma_{x_n}$. Let
$\sum_k l_{k}  |e_{k}\rangle\langle e_{k}|$ be a
spectral decomposition  of $P\varsigma:=\sum_{x^n\in{\mathcal{X}'}^n}P^n(x^n)\varsigma_{x^n}$, where $l_{k} \in
\mathbb{R}^+$, $\sum_{k}l_{k}=1$. For $\alpha > 0$ denote
$G_{\alpha} := \{k
:2^{-n[S(\sum_{x^n\in{\mathcal{X}'}^n}P^n(x^n)\varsigma_{x^n})-\alpha]}
\leq l_{k}\leq
2^{-n[S(\sum_{x^n\in{\mathcal{X}'}^n}P^n(x^n)\varsigma_{x^n})+\alpha]}\}$.
Denote $\Pi_{\varsigma, \alpha} := \sum_{k \in G_{\alpha}}
|e_k\rangle\langle e_k|$. Then $\Pi_{\varsigma ,\alpha}$ commuting with
$P\varsigma $ and satisfying
\[ \mathrm{tr} \left( P\varsigma
 \Pi_{P\varsigma ,\alpha} \right) \geq 1-\frac{d}{4n\alpha ^2}\text{ ,}\]
\[ \mathrm{tr} \left( \Pi_{\rho ,\alpha} \right)
 \leq 2^{ S(\sum_{x^n\in{\mathcal{X}'}^n}P(x^n)\varsigma_{x^n}) + Kd\alpha \sqrt{n}}\text{ ,}\]
\[  \Pi_{P\varsigma ,\alpha} \cdot P\varsigma \cdot \Pi_{P\varsigma ,\alpha} \leq
2^{ -S(\sum_{x^n\in{\mathcal{X}'}^n}P(x^n)\varsigma_{x^n}) + Kd\alpha
\sqrt{n}}\Pi_{P\varsigma ,\alpha}\text{ ,}\]
\[  \mathrm{tr} \left( P\varsigma \cdot \Pi_{P\varsigma, \alpha \sqrt{a}} \right)
 \geq 1-\frac{ad}{4n\alpha ^2}\text{ ,}\] where  $a := \#\mathcal{X}'$ and
 $K$ is a positive constant.\vspace{0.15cm}

Let $\sum_{k} l_{x^n,k}
|e_{x^n,j}\rangle\langle e_{x^n,k}|$ be a spectral
decomposition  of $\varsigma_{x^n}$,  where $l_{x^n,k}
\in \mathbb{R}^+$, $\sum_{k}l_{x^n,k}=1$.
For $\alpha >0$ denote $G_{x^n,\alpha} := \{k
:2^{-n[S(\varsigma_{x^n})-\alpha]} \leq l_{x^n,k}\leq
2^{-n[S(\varsigma_{x^n})+\alpha]}\}$, and $\Pi_{\varsigma_{x^n},\alpha} := \sum_{k
\in G_{x^n,\alpha}} |e_{x^n,k}\rangle\langle e_{x^n,k}|$.

The subspace projector $\Pi_{\varsigma_{x^n},\alpha}$ commutes with $\varsigma_{x^n}$
and satisfies:
\[ \mathrm{tr} \left( \varsigma_{x^n} \Pi_{\varsigma_{x^n},\alpha} \right)
 \geq 1-\frac{ad}{4n\alpha ^2}\text{ ,}\]
\[ \mathrm{tr} \left( \Pi_{\varsigma_{x^n},\alpha} \right)
\leq 2^{\sum_{x^n\in{\mathcal{X}'}^n}P(x^n)S(\varsigma_{x^n})  +
Kad\alpha \sqrt{n}}\text { ,}\]
\begin{align*}   &  \Pi_{\varsigma_{x^n},\alpha} \cdot \varsigma_{x^n} \cdot \Pi_{\varsigma_{x^n},\alpha} \notag\\
&\leq 2^{ -\sum_{x^n\in{\mathcal{X}'}^n}P(x^n)S(\varsigma_{x^n}) +
Kad\alpha \sqrt{n}}\Pi_{\varsigma_{x^n},\alpha}\text{ ,}\end{align*}

where $K$ is a positive constant.\vspace{0.15cm}

\section{Main Result}

\begin{theorem} Let $\mathcal{W} := \{(\rho_{x,t}, \sigma_{x,t}) : x\in
\mathcal{X}, t \in \Theta\}$ be a classical-quantum arbitrarily
varying wiretap channel, if for all $t\in\Theta$ it holds: $\{\rho_{x,t}, x\in\mathcal{X}\}$ is not
symmetrizable, then the largest achievable secrecy rate, called secrecy capacity, of
$\mathcal{W}$, is bounded as follow,
\begin{equation}C(\mathcal{W}) \geq \max_{P\in\mathcal{P}}
\left(\min_{Q\in\mathcal{Q}}\chi\left(P,\{\rho^Q_x: x\in\mathcal{X}\}\right) -
\lim_{n\rightarrow\infty}\max_{t^n\in\Theta^n}\frac{1}{n}\chi(P^n,\{\sigma_{x^n,t^n}: x^n\in\mathcal{X}^n\})\right)\text{
,}\label{t1}\end{equation} where $\mathcal{P}$ are distributions on
$\mathcal{X}$, $\mathcal{Q}$ are distributions on $\Theta$, and
$\rho^Q_x = \sum_{t\in\Theta}Q(t)\rho_{x,t}$ for
$Q\in\mathcal{Q}$.\\[0.15cm]
If $\{\rho_{x,t}, \in
\mathcal{X}, t \in \Theta\}$ is not symmetrizable, then the secrecy capacity of  $\mathcal{W}$ with CSI at the transmitter is  bounded as follow
 \begin{align}&C_{CSI}(\mathcal{W}) \geq \min_{Q\in\mathcal{Q},t^n\in\Theta^n} \max_{P\in\mathcal{P}}
\left(\chi(P,\{\rho^Q_x: x\in\mathcal{X}\}) -
\lim_{n\rightarrow\infty}\frac{1}{n}\chi(P^n,\{\sigma_{x^n,t^n}: x^n\in\mathcal{X}^n\})\right)\text{ .}\label{t2}\end{align}
Is $\{\rho_{x,t}, x\in\mathcal{X}\}$ symmetrizable for some $t\in\Theta$, then
we have:
\begin{equation}C(\mathcal{W}) = C_{CSI}(\mathcal{W}) =0\text{
.}\label{t3}\end{equation}
\end{theorem}\vspace{0.3cm}

\begin{proof}

At first, we are going to prove (\ref{t1}).\vspace{0.15cm}

For $P\in\mathcal{P}$
denote
$\rho^{P,Q}:=\sum_{x\in\mathcal{X}}P(x)\rho_{x}^{Q}$. Let \[J_n
=\left\lfloor2^{n\min_{Q}\chi\left(P,\{\rho^Q_x:
x\in\mathcal{X}\}\right) -
\max_{t^n\in\Theta^n}\chi\left(P^n,\{\sigma_{x^n,t^n}:
x^n\in\mathcal{X}^n\}-2n\eta\right)}\right\rfloor\text{ ,}\]
\[L_n=\left\lfloor2^{\max_{t^n\in\Theta^n}\chi\left(P^n,\{\sigma_{x^n,t^n}: x^n\in\mathcal{X}^n\}+n\eta\right)}\right\rfloor\text{ ,}\]
where $\eta$ is a
 positive constant.\vspace{0.2cm}

  Let
${P'} (x^n):= \begin{cases} \frac{P^{ n}(x^n)}{P^{ n}
(\mathcal{T}^n_{P,\delta})} & \text{if } x^n \in \mathcal{T}^n_{P,\delta}\\
0 & \text{else} \end{cases}\text{ ,}$ and $X^n :=
\left(X_{j,l}^n\right)_{j \in \{1, \cdots, J_n\}, l \in \{1, \cdots,
L_{n}\}}$ be a family of random matrices such that their entries are
i.i.d. according to $P'$.\vspace{0.3cm}

Fix $P\in\mathcal{P}$. Denote $Q^n=(Q_1,\cdots,Q_n)\in\mathcal{Q}$.
Let $\rho^{P,Q^n} :=
\rho^{P,Q_1}\otimes\rho^{P,Q_2}\otimes\cdots\otimes\rho^{P,Q_n}$,
and $\sum_{j^n} \lambda_{j^n}^{P,Q^n} |e_{j^n}^{P,Q^n}\rangle\langle
e_{j^n}^{P,Q^n}|$ be a spectral decomposition  of $\rho^{P,Q^n}$,
where $\lambda_{j^n}^{P,Q^n} \in \mathbb{R}^+$,
$\sum_{j^n}\lambda_{j^n}^{P,Q^n}=1$.

For $\delta >0$ denote $F_{\delta,Q^n} := \{j^n
:2^{-\sum_{i=1}^{n}H(\rho^{P,Q_i})-n\delta}
\leq\lambda_{j^n}^{P,Q^n}\leq
2^{-\sum_{i=1}^{n}H(\rho^{P,Q_i})+n\delta}\}$, and
$\Pi_{\delta}^{P,Q^n} := \sum_{j^n \in F_{\delta,Q^n}}
|e_j^{P,Q^n}\rangle\langle e_j^{P,Q^n}|$.\vspace{0.15cm}

For any $x^n \in \mathcal{X}^n$ let $\rho_{x^n}^{Q^n}
:=\rho_{x_1}^{Q_1}\otimes\rho_{x_2}^{Q_2}\otimes\cdots\otimes\rho_{x_n}^{Q_n}$.
 Let
$ \sum_{j^n} \lambda_{x^n,j^n}^{Q^n}
|e_{x^n,j^n}^{Q^n}\rangle\langle e_{x^n,j^n}^{Q^n}|$ be a spectral
decomposition  of $\rho_{x^n}^{Q^n}$, where $\lambda_{x^n,j^n}^{Q^n}
\in \mathbb{R}^+$, $\sum_{j^n}\lambda_{x^n,j^n}^{Q^n}=1$.

For $\delta >0$ denote $F_{x^n,Q^n,\delta} := \{j_n
:2^{-\sum_{i=1}^{n}\sum_{x\in\mathcal{X}}P(x)H(\rho_{x}^{Q_i})-n\delta}
\leq\lambda_{x^n,j}^{Q^n}\leq
2^{-\sum_{i=1}^{n}\sum_{x\in\mathcal{X}}P(x)H(\rho_{x}^{Q_i})+n\delta}\}$,
and $\Pi_{x^n,\delta}^{Q^n} := \sum_{j_n \in F_{x^n,Q^n,\delta}}
|e_{x^n,j}^{Q^n}\rangle\langle e_{x^n,j}^{Q^n}|$.\vspace{0.3cm}

Our proof bases on the following two lemmas. The first lemma is due to
Rudolf Ahlswede and Vladimir Blinovsky, the second one (the Covering Lemma)
is due to Rudolf Ahlswede and
Andreas Winter.

\begin{lemma}[cf. \cite{Ahl/Bli}]\label{l1}

Let  $\{\varrho_{x,t}, x\in\mathcal{X},t\in\Theta \}$ be a classical-quantum
 arbitrarily varying wiretap channel, defined in sense of \cite{Ahl/Bli},
 where
$\mathcal{X}$ is the set of code symbols and $\Theta$ is the set of
states of the classical-quantum
 arbitrarily varying wiretap channel.

For $\{c_i^n :i=1,\cdots,N\}\subset \mathcal{X}^n$ and distribution
$Q^n=(Q_1,Q_2,\cdots Q_n)$ on $\Theta^n$  define
\begin{equation}D_i^{Q^n} := \left(\sum_{j=1}^{N} \Pi_{\delta}^{P,Q^n}\Pi_{c_{j}^n,\delta}^{P,Q^n}
\Pi_{\delta}^{P,Q^n}\right)^{-1/2}
\Pi_{\delta}^{P,Q^n}\Pi_{c_{i}^n,\delta}^{P,Q^n}
\Pi_{\delta}^{P,Q^n}
\left(\sum_{j=1}^{N}\Pi_{\delta}^{P,Q^n}\Pi_{c_{j}^n,\delta}^{P,Q^n}
\Pi_{\delta}^{P,Q^n}\right)^{-1/2} \text{ .}\end{equation}

Define the the set of the  quantum codes
\begin{equation}\mathcal{C}:=\left\{ C^{Q^n}=\{(c_i^n,D_i^{Q^n}):
i=1,\cdots,N\}:c_i^n \in \mathcal{X}^n\forall i, Q^n \text{ is a
distribution on} \Theta^n \right\}\text{ ,}\end{equation}

If $\frac{\log N}{n} <
\min_{Q}\chi\left(P,\{\varrho^Q_x:x\in\mathcal{X}\}\right)-\delta$,
where $\delta$ is a positive constant, and assume $\{\varrho_{x,t},
x\in\mathcal{X},t\in\Theta\}$ is not symmetrizable. then following
holds. For any $\epsilon >0$, if $n$ is large enough, then  there
exist a distribution $G$ on $\mathcal{C}$ such that
\begin{equation}\max_{t^n\in\Theta^n} \sum_{C\in\mathcal{C}}P_e(C, t^n)G(C) < \epsilon\text{ .}\end{equation}\vspace{0.15cm}

\end{lemma}

\begin{lemma}[Covering Lemma, cf. \cite{Wil}] Suppose we are given a finite set $\mathcal{Y}$,
 an ensemble $\{\sigma_y : y\in \mathcal{Y}\}$ with probability distribution $p_\mathcal{Y}$ on $\mathcal{Y}$.
Suppose there exist a total subspace projector $\Pi$ and codeword subspace projectors $\{\Pi_{y}\}_{y\in \mathcal{Y}}$ ,
they project onto subspaces of the Hilbert space in which the states $\{\sigma_y\}$ exist, and these projectors and the ensemble
satisfy the following conditions:

\[\mathrm{tr} \left(\sigma_y\Pi\right)\geq 1-\epsilon\]
\[\mathrm{tr} \left(\sigma_y\Pi_y\right)\geq 1-\epsilon\]
\[\mathrm{tr} \left(\Pi\right)\leq c\]
\[\Pi_y\sigma_y\Pi_y\leq \frac{1}{d}\Pi_y\]

Suppose that $\mathcal{M}\subset\mathcal{Y}$ is a set of size $|\mathcal{M}|$ with elements $\{m\}$, $C = \{C_m\}_{m \in \mathcal{M}}$
is a random code  where
the codewords $C_m$ are chosen according to the distribution $p_\mathcal{Y}
(y)$, and an ensemble $\{\sigma_{C_m} : m\in \mathcal{M}\}$ with uniform
distribution on $\mathcal{M}$, then

\begin{equation}Pr\left\{\left\|\sum_{y \in \mathcal{Y}} p_\mathcal{Y} (y)  \sigma_y -
\frac{1}{|\mathcal{M}|}\sum_{m\in \mathcal{M}} \sigma_{C_m} \right\|_1 \leq \epsilon + 4\sqrt{\epsilon} +24\sqrt[4]{\epsilon}\right\} \geq 1-2c\exp
\left(-\frac{\epsilon^3 |\mathcal{M}| d}{2 \log 2 c}\right)\text{ .}\end{equation}

\end{lemma}\vspace{0.3cm}

 Let $
\{X_{j,l}^n\}_{j \in \{1, \cdots, J_n\}, l \in \{1, \cdots,
L_{n}\}}$ be a family of random matrices such that  the entries of
$\{X_{j,l}^n\}_{j \in \{1, \cdots, J_n\}, l \in \{1, \cdots,
L_{n}\}}$ are i.i.d. according to ${P'}^n$.\vspace{0.2cm}

For every realization $\left(x_{j,l}^n\right)_{j \in \{1, \cdots,
J_n\}, l \in \{1, \cdots, L_{n}\}}$ of $\left(X_{j,l}^n\right)_{j
\in \{1, \cdots, J_n\}, l \in \{1, \cdots, L_{n}\}}$
 and distribution $Q^n$ on $\Theta^n$  define

\begin{equation}D_{x_{j,l}^n}^{Q^n} := \left(\sum_{j'=1}^{J_n}\sum_{l'=1}^{L_n} \Pi_{\delta}^{P,Q^n}\Pi_{x_{j',l'}^n,\delta}^{P,Q^n}
\Pi_{\delta}^{P,Q^n}\right)^{-1/2}
\Pi_{\delta}^{P,Q^n}\Pi_{x_{j,l}^n,\delta}^{P,Q^n}
\Pi_{\delta}^{P,Q^n}
\left(\sum_{j'=1}^{J_n}\sum_{l'=1}^{L_n}\Pi_{\delta}^{P,Q^n}\Pi_{x_{j',l'}^n,\delta}^{P,Q^n}
\Pi_{\delta}^{P,Q^n}\right)^{-1/2} \text{ .}\end{equation}
\vspace{0.15cm}

Let
$\sum_k \lambda_{k}^{t^n}  |e_{k}^{t^n}\rangle\langle e_{k}^{t^n}|$ be a
spectral decomposition  of $P\rho_{t^n}:=\sum_{x^n\in{\mathcal{X}}^n}P^n(x^n)\rho_{x^n,t^n}$
We denote $G_{\alpha}^{t^n} := \{k
:2^{-n[H(\sum_{x^n\in\mathcal{X}^n}P^n(x^n)\rho_{x^n,t^n})-\alpha]}
\leq\lambda_{k}^{t^n}\leq
2^{-n[H(\sum_{x^n\in\mathcal{X}^n}P^n(x^n)\rho_{x^n,t^n})+\alpha]}\}$
and $\Pi_{\rho_{t^n}, \alpha} := \sum_{k \in G_{\alpha}^{t^n}}
|e_j^{t^n}\rangle\langle e_k^{t^n}|$,  where $\rho_{x^n,t^n}:=
\rho_{x_1,t_1}\otimes\cdots\otimes\rho_{x_n,t_n}$.
Let $\sum_{k} \lambda_{x^n,k}^{t^n}
|e_{x^n,k}^{t^n}\rangle\langle e_{x^n,k}^{t^n}|$ be a spectral
decomposition  of $\rho_{x^n,t^n}$.
Denote $G_{x^n,\alpha}^{t^n} := \{k
:2^{-n[H(\rho_{x^n,t^n})-\alpha]} \leq\lambda_{x^n,k}^{t^n}\leq
2^{-n[H(\rho_{x^n,t^n})+\alpha]}\}$, and $\Pi_{\rho_{x^n,t^n},\alpha} := \sum_{k
\in G_{x^n,\alpha}^{t^n}} |e_{x^n,k}^{t^n}\rangle\langle e_{x^n,k}^{t^n}|$.

Define
\begin{equation}\overline{\sigma}_{x^n,t^n} := \Pi_{P\rho_{t^n}, \alpha \sqrt{a}}\Pi_{\rho_{x^n, t^n},\alpha}
 \cdot \sigma_{x^n,t^n} \cdot \Pi_{\rho_{x^n, t^n},\alpha}\Pi_{P\rho_{t^n}, \alpha \sqrt{a}}\text{ .}\end{equation}

By
\begin{lemma}[cf.  \cite{Win}] \label{l2}
Let $\rho$ be a state and $X$ be a positive operator with $X  \leq
id$ (the identity  matrix) and $1 - \mathrm{tr}(\rho X)  \leq
\lambda \leq1$. Then
\[ \| \rho -\sqrt{X}\rho \sqrt{X}\|_1 \leq \sqrt{8\lambda}\text{ .}
\] \end{lemma}

  and the fact
that $\Pi_{P\rho_{t^n}, \alpha \sqrt{a}}$ and $\Pi_{\rho_{x^n, t^n},\alpha}$ are
both projection matrices,  for any $t^n$ and $x^n$ it holds:
\[\|\overline{\sigma}_{x^n,t^n}-\sigma_{x^n,t^n} \|_1  \leq
\sqrt{\frac{2(ad+d)}{n\alpha^2}} \text{ ,}\] Thus for any positive $\alpha$  and any
positive $\eta$ if $n$ is large enough
\begin{equation}\|\overline{\sigma}_{x^n,t^n}-\sigma_{x^n,t^n} \|_1  \leq
\eta \text{ .}\label{e2}\end{equation}

Since \[\mathrm{tr} \left( \Pi_{P\rho_{t^n}, \alpha \sqrt{a}}
\right)
 \leq 2^{ S(\sum_{x^n\in\mathcal{X}^n}P^n(x^n)\rho_{x^n,t^n})}\text{ ,}\] \[\Pi_{\rho_{x^n, t^n},\alpha}
 \cdot \sigma_{x^n,t^n} \cdot \Pi_{\rho_{x^n, t^n},\alpha}
\leq 2^{
-\sum_{x^n\in\mathcal{X}^n}P^n(x^n)S(\rho_{x^n,t^n})}\Pi_{\rho_{x^n, t^n},\alpha}\text{
,}\]and
\[L_n \geq
2^{\chi\left(P^n,\{\sigma_{x^n,t^n}: x^n\in\mathcal{X}^n\})\right)+2n\delta}=
\frac{2^{\sum_{x^n}P^n(x^n)S(\rho_{x^n,t^n})+n\delta}}{2^{S\left(\sum_{x^n}P^n(x^n)\rho_{x^n,t^n}\right)-n\delta}}\text{
,}\] by applying covering lemma, for every $t^n$ and $j'\in\{1,\cdots,J_n\}$ there is a positive constant $c_1'$ such that for any $\nu>0$,
\begin{equation} Pr\left\{\left\| \frac{1}{L_n} \frac{1}{J_n} \sum_{l=1}^{L_n} \sum_{j=1}^{J_n} \overline{\sigma}_{X_{j,l}^n,t^n}   -
\frac{1}{L_n} \sum_{l=1}^{L_n} \overline{\sigma}_{X_{j',l}^n,t^n} \right\|_1  <\nu
\right\} \geq 1- 2^{-\nu^3 2^{nc_1'}}
  \label{e1+}\text{ .}\end{equation}
Since $|\Theta^n|=O(2^n)$, and $J_n \ll 2^{\nu^3 2^{nc_1'}}$, there
is a positive constant $c_1$ such that for any $\nu>0$,
\begin{equation} Pr\left\{\left\| \frac{1}{L_n} \frac{1}{J_n} \sum_{l=1}^{L_n} \sum_{j=1}^{J_n} \overline{\sigma}_{X_{j,l}^n,t^n}   -
\frac{1}{L_n} \sum_{l=1}^{L_n} \overline{\sigma}_{X_{j',l}^n,t^n} \right\|_1  <\nu
 \text{ }\forall j' \in \{1,\cdots,J_n\}\forall t^n \in \Theta^n \right\} \geq 1- 2^{-\nu^3 nc_1}
  \label{e1}\text{ .}\end{equation}

Denote the set of all codes
$\left\{\left(x_{j,l}^{,n},D_{x_{j,l}^{n}}^{Q^n}\right): j =1,
\cdots, J_n, l =1, \cdots, L_{n}\right\}$, where
  $(x_{j,l})_{j=1,\cdots,J_n, l=1,\cdots,L_n}$ are realizations of
$(X_{j,l})_{j=1,\cdots,J_n, l=1,\cdots,L_n}$, such that
\[\left\| \frac{1}{L_n} \frac{1}{J_n} \sum_{l=1}^{L_n} \sum_{j=1}^{J_n} \overline{\sigma}_{x_{j,l}^n,t^n}   -
\frac{1}{L_n} \sum_{l=1}^{L_n} \overline{\sigma}_{x_{j',l}^n,t^n} \right\|_1  <\nu
 \text{ }\forall j' \in \{1,\cdots,J_n\}\forall t^n \in \Theta^n \] by $\mathcal{C}_{\nu}'$.\vspace{0.2cm}

Now we want to show the following alternative result to Lemma \ref{l1}.\vspace{0.15cm}

If $n$ is large enough then for any  any positive $\nu$, there exist a distribution $G$ on
$\mathcal{C}_{\nu}'$ such that
\begin{equation}\max_{t^n\in\Theta^n} \sum_{C\in\mathcal{C}_{\nu}'}P_e(C, t^n)G(C) < \epsilon\text{ .}\label{altn}\end{equation}
\vspace{0.15cm} In  \cite{Ahl/Bli}, following inequality is shown.
There is a positive constant $c_2$ such that for any positive $\nu$
\begin{equation}Pr\left\{1-\frac{1}{L_n} \frac{1}{J_n} \sum_{l=1}^{L_n} \sum_{j=1}^{J_n}
\mathrm{tr} \left(D_i^{Q^n}\rho_{X_{j,l}^n}^{Q^n}\right)\leq \nu \right\}\geq 1- J_nL_n\cdot 2^{n[\min_Q(H(\rho^{P,Q})-\sum_{x}P(x)H(\rho^{Q}_x))-c_2]}\text{ ,}\end{equation}
where $c_2$ is some positive constant. Since
\[J_nL_n\leq 2^{n\left[\min_{t^n\in\Theta^n}\chi(P,\{\rho_{x^n,t^n}:
x^n\in\mathcal{X}^n\})
-\eta\right]}\text{ ,}\]
There is a  positive constant $c_3$ such that if $n$ is large enough then
\[ J_nL_n\cdot 2^{n[\min_Q(H(\rho^{P,Q})-\sum_{x}P(x)H(\rho^{Q}_x))-c_2]} \leq 2^{-nc_3}\]

Denote  the set of all codes
$\left\{\left(x_{j,l}^{,n},D_{x_{j,l}^{n}}^{Q^n}\right): j =1,
\cdots, J_n, l =1, \cdots, L_{n}\right\}$ such that
\[1-\frac{1}{L_n} \frac{1}{J_n} \sum_{l=1}^{L_n} \sum_{j=1}^{J_n}
\mathrm{tr} \left(D_i^{Q^n}\rho_{x_{j,l}^n}^{Q^n}\right)\leq \nu
\] by $\mathcal{C}_{\nu}''$.\vspace{0.2cm}

We have \[Pr(\mathcal{C}_{\nu}'\cap\mathcal{C}_{\nu}'')\geq 1- 2^{-n\nu c_1} - 2^{-nc_3}\text{ ,}\]
therefore  if $n$ is large enough, $\mathcal{C}_{\nu}'\cap\mathcal{C}_{\nu}''$ is not empty.
This means if  is large enough,
then for any positive $\nu$ and for each set
of distributions $T^n = (T_1,\cdots T_n)$ on $\Theta^n$,  there exists a
$C^{X_{T^n}}\in\mathcal{C}_{\nu}'$ with a positive probability
such that,
\[\sum_{t^n\in\Theta^n} T^n(t^n) P_e(C^{X_{T^n}},t^n)\leq \nu\text{ ,}\]
where $T^n(t^n)=T_1(t_1) T_2(t_2)\cdots T_n(t_n)$.\vspace{0.15cm}

Let us denote the set of distributions on $\mathcal{C}_{\nu}'$ by $\Omega_{\mathcal{C}_{\nu}'}$.
By applying the minimax theorem
for mixed strategies (cf. \cite{Ahl/Bli}), we have
\[\max_{T^n}\min_{G\in\Omega_{\mathcal{C}_{\nu}'}}\sum_{t^n\in\Theta^n, C\in\mathcal{C}_{\nu}'}
 T^n(t^n)G(C)P_e(C,t^n)=\min_{G\in\Omega_{\mathcal{C}_{\nu}'}}\max_{T^n}\sum_{t^n\in\Theta^n, C\in\mathcal{C}_{\nu}'}
 T^n(t^n)G(C)P_e(C,t^n)\text{ .}\]
Therefore (\ref{altn}) holds.\vspace{0.3cm}

Now we are going to  use the derandomization technique in  \cite{Ahl/Bli} to
build a deterministic code.\vspace{0.15cm}

Consider now $n^2$ independent and
identically distributed  random variables
$Z_1,Z_2,\cdots,Z_{n^2}$ with values in  $\mathcal{C}_{\nu}'$
such that $P(Z_i=C)=G(C)$ for all $ C \in\mathcal{C}_{\nu}'$ and
for all  $i\in\{1,\cdots,n^2\}$. Then for given $t^n\in\Theta^n$
\begin{equation}G\left(\sum_{i=1}^{n^2} P_e(Z_i,t^n)>\lambda n^2\right)<e^{-\lambda n^2}
\text{ ,}\label{n^2}\end{equation} where $\lambda:=\log (\nu\cdot
e^2 +1)$. If $n$ is large enough then $1-e^{-\lambda n^2}$ is
positive, this means $\left\{C^{z_i} \text{ is a realization of }
Z_i : \sum_{i=1}^{n^2} P_e(C^{z_i},t^n)<\lambda n^2\right\}$ is not
the empty set, since $G(\emptyset)=0$ by the definition of
distribution.

 In \cite{Ahl/Bli}, it is shown that if $\left\{C^{z_i} \text{ is a realization of } Z_i :
\sum_{i=1}^{n^2} P_e(C^{z_i},t^n)<\lambda n^2\right\}$ is not the empty set,
there
exist  codes $C_1,C_2,\cdots,C_{n^2}\in \mathcal{C}_{\nu}'$, where we denote $C_i=
\left\{\left(x_{j,l}^{(i),n},D_{x_{j,l}^{(i),n}}^{Q^n}\right): j
=1, \cdots, J_n, l =1, \cdots,
L_{n}\right\}$ for $i\in\{1,\cdots,n^2\}$ with a positive probability such that
\begin{equation} \frac{1}{n^2}\sum_{i=1}^{n^2} P_e(C_i,t^n)<\lambda\text{ .}\label{n2'}\end{equation}
\vspace{0.2cm}

Following fact is trivial. There is a code $\left\{(c^{\mu(n)}_i,D_i): i=1,\cdots,n^2\right\}$ of length $\mu(n)$, where
$\mu(n)=o(n)$ (this code does not need to be secure against the wiretapper, i.e. we allow the wiretapper
to have the full knowledge of $i$),
such that for any positive $\vartheta$ if $n$ is large enough then
\[\min_{t^n\in\Theta^n} \frac{1}{n^2} \sum_{i=1}^{n^2} \mathrm{tr}(\rho_{c_i^{\mu(n)},t^n} D_i)\geq 1-\vartheta\text{ .}\]

By (\ref{n2'})
 we can construct a code of length $\mu(n)+n$ (cf. \cite{Ahl/Bli})
\[C^{det}=\left\{\left(c^{\mu(n)}_i\otimes x_{j,l}^{(i),n},  D_i\otimes
D_{x_{j,l}^{(i),n}}^{Q^n} \right):i=1,\cdots,n^2, j=1,\cdots,J_n,
l=1,\cdots,L_n\right\}\text{ ,}\] which is a juxtaposition of words
of the  code $\{(c^{\mu(n)}_i, D_i): i=1,\cdots,n^2\}$ and the words
of  code $C_i=\{(x_{j,l}^{(i),n}
D_{x_{j,l}^{(i),n}}^{Q^n}):j=1,\cdots,J_n, l=1,\cdots,L_{n}\}$,
 with following feature.
$C^{det}$ is a deterministic code  with $n^2J_nL_n$ codewords such
that for any positive $\epsilon$ if $n$ is large enough then

\begin{equation}\max_{t^n \in \Theta^n} P_e(C^{det}, t^n) < \epsilon\text{ .}\label{q1*}\end{equation}

Furthermore, since $C_1,C_2,\cdots,C_{n^2}\in \mathcal{C}_{\nu}'$, for all $i\in\{1,\cdots,n^2\}$,
$t^n\in\Theta^n$, and $j' \in \{1,\cdots,J_n\}$ we have

\begin{equation}\| \frac{1}{L_n} \sum_{l=1}^{L_n} \sigma_{x_{j',l}^{(i),n},t^n} -
  \frac{1}{L_n} \frac{1}{J_n} \sum_{l=1}^{L_n} \sum_{j=1}^{J_n} \sigma_{x_{j,l}^{(i),n},t^n} \|_1  <3\eta\text{ .}\end{equation}\vspace{0.2cm}

\begin{lemma}[Fannes inequality, cf. \cite{Win}]\label{fann}
Let $\mathfrak{X}$ and $\mathfrak{Y}$ be two states in a
$d$-dimensional complex Hilbert space and
$\|\mathfrak{X}-\mathfrak{Y}\|_1 \leq \mu < \frac{1}{e}$, then
\[|S(\mathfrak{X})-S(\mathfrak{Y})| \leq \mu \log d
- \mu \log \mu \text{ .}\]\end{lemma}

Let $W$ be a random variable  uniformly distributed on
$\{1,\cdots,J_n\}$, by Lemma \ref{fann}, for all $t^n \in \Theta^n$ and all $i\in \{1,\cdots,n^2\}$
\begin{align*} &
\chi\left(W,\{\frac{1}{L_n}\sum_{l=1}^{L_n}\sigma_{x_{j,l}^{(i),n},t^n}
: j =1 \cdots J_n\}\right)\\
&= S\left(  \frac{1}{J_n}\frac{1}{L_n}\sum_{j=1}^{J_n}
\sum_{l=1}^{L_n}\sigma_{x_{j,l}^{(i),n},t^n}\right)
-\frac{1}{J_n}\sum_{j=1}^{J_n} S\left( \frac{1}{L_n}\sum_{l=1}^{L_n}\sigma_{x_{j,l}^{(i),n},t^n}\right)\\
&= \frac{1}{J_n} \sum_{j'=1}^{J_n} \left[S\left(
\frac{1}{J_n}\frac{1}{L_n} \sum_{j=1}^{J_n} \sum_{l=1}^{L_n}
\sigma_{x_{j,l}^{(i),n},t^n}\right)- S\left(\frac{1}{L_n}\sum_{l=1}^{L_n} \sigma_{x_{j',l}^{(i),n},t^n}\right)\right]\\
&\leq 3\eta \log d -3\eta \log 3\eta\text{ .}
\end{align*}

Therefore for any $\zeta>0$ we can choose such $\eta$ that for all $i\in \{1,\cdots,n^2\}$
(i.e. even when the wiretapper
has the full knowledge of $i$), for all $t^n\in\Theta^n$,
\begin{equation}\chi\left(W,\left\{\left[\frac{1}{L_n}\sum_{l=1}^{L_n}\sigma_{x_{j,l}^{(i),n},t^n}\right]
: j =1 \cdots J_n\right\}\right) < \zeta\label{q2}\text{ .}
\end{equation}

By (\ref{q1*}) and (\ref{q2}),  we see that for any distribution $P$
on $\mathcal{X}$ and any positive $\delta$, we can find a $(n,\epsilon)$-code with
secrecy rate $\min_{Q}\chi\left(P,\{\rho^Q_x:
x\in\mathcal{X}\}\right)  -
\frac{1}{n}\max_{t^n\in\Theta^n}\chi\left(P^n,\{\sigma_{x^n,t^n}:
x^n\in\mathcal{X}^n\}\right) - \delta$. Therefore (\ref{t1})
follows.\vspace{0.4cm}

Now, we are going to prove (\ref{t2}).\vspace{0.15cm}

Fix $P$, let $J_n =\left\lfloor 2^{\min_{Q\in\mathcal{Q},t^n\in\Theta^n}\left[
n\chi(P,\{\rho^Q_x: x\in\mathcal{X}\}) -
\chi(P^n,\{\sigma_{x^n,t^n}: x^n\in\mathcal{X}^n\})\right]-2n\eta}\right\rfloor$,
$L_{t^n}=\left\lfloor2^{\chi(P,\{\sigma_{x^n,t^n}:
x^n\in\mathcal{X}^n\})+n\eta}\right\rfloor$, where $\eta$ is a
 positive constant.  For any $t^n\in\Theta^n$ let $X^{(t^n)}:=
\{X_{j,l}^{(t^n)}\}_{j \in \{1, \cdots, J_n\}, l \in \{1, \cdots,
L_{t^n}\}}$ be a family of random matrices whose components are
i.i.d. according to ${P'}$.\vspace{0.2cm}

 For any realization
$(x^{(t^n)}_{j,l})_{j \in \{1, \cdots, J_n\}, l \in \{1, \cdots,
L_{t^n}\}}$ of  $(X^{(t^n)}_{j,l})_{j \in \{1, \cdots, J_n\}, l \in
\{1, \cdots, L_{t^n}\}}$ and distribution $Q^n$ on $\Theta^n$ define

\begin{align}&D_{x_{j,l}^{(t^n)}}^{Q^n,t^n}\notag\\
& := \left(\sum_{j'=1}^{J_n}\sum_{l'=1}^{L_{t^n}}
\Pi_{\delta}^{P,Q^n}\Pi_{{x_{j',l'}^{(t^n)}},\delta}^{P,Q^n}
\Pi_{\delta}^{P,Q^n}\right)^{-1/2}\notag\\
&\cdot\Pi_{\delta}^{P,Q^n}\Pi_{{x_{j,l}^{(t^n)}},\delta}^{P,Q^n}
\Pi_{\delta}^{P,Q^n}\notag\\
&\cdot\left(\sum_{j'=1}^{J_n}\sum_{l'=1}^{L_{t^n}}\Pi_{\delta}^{P,Q^n}\Pi_{{x_{j',l'}^{(t^n)}},\delta}^{P,Q^n}
\Pi_{\delta}^{P,Q^n}\right)^{-1/2} \text{ .}\end{align}

Since $L_{t^n} \geq 2^{\chi\left(P^n,\{\sigma_{x^n,t^n}:
x^n\in\mathcal{X}^n\}\right)+2n\delta}=
\frac{2^{\sum_{x^n}P^n(x^n)S(\rho_{x^n,t^n})+n\delta}}{2^{S\left(\sum_{x^n}P^n(x^n)\rho_{x^n,t^n}\right)-n\delta}}$,
by applying covering lemma, there is a positive $c_1'$ such that for any positive $\eta$ we have:
\begin{equation} Pr\left\{\left\| \frac{1}{L_{t^n}} \frac{1}{J_n}
 \sum_{l=1}^{L_{t^n}} \sum_{j=1}^{J_n} \overline{\sigma}_{X_{j,l}^{(t^n)},t^n}   -
\frac{1}{L_{t^n}} \sum_{l=1}^{L_{t^n}}
\overline{\sigma}_{X_{j',l}^{(t^n)}, t^n}\right\|_1 <\eta  \text{
}\forall j' \in \{1,\cdots,J_n\}
\right\} = 1-2^{-\nu^3 c_1''}
  \label{e1'}\end{equation}

Denote all $\{(x_{j,l}^{(t^n)},
D_{x_{j,l}^{(t^n)}}^{Q^n}):j=1,\cdots,J_n, l=1,\cdots,L_{t^n}\}$
such that
\[\left\| \frac{1}{L_{t^n}} \frac{1}{J_n}
 \sum_{l=1}^{L_{t^n}} \sum_{j=1}^{J_n} \overline{\sigma}_{x_{j,l}^{(t^n)},t^n}   -
\frac{1}{L_{t^n}} \sum_{l=1}^{L_{t^n}}
\overline{\sigma}_{x_{j',l}^{(t^n)}, t^n}\right\|_1 <\eta \forall j'
\in \{1,\cdots,J_n\}\text{ ,}\] by $\mathcal{C}_{\nu
}^{t^n}$.\vspace{0.2cm}

Since   $\frac{1}{n}\log (J_n\cdot
L_{t^n}) \leq \min_{Q}\chi\left(P,\{\rho^Q_x:
x\in\mathcal{X}\}\right)-2\delta$, analogue to our proof for
(\ref{t1}), if $n$ is large enough then there are $n^2$
codes$\left\{\{(x_{j,l}^{(i),(t^n)},
D_{x_{j,l}^{(i),(t^n)}}^{Q^n}):j=1,\cdots,J_n,
l=1,\cdots,L_{t^n}\}:i=1,\cdots,n^2\right\}$ $ \in
\mathcal{C}_{\nu}^{t^n}$ such that we
 can construct a code $C^{det}_{t^n}$ which is a juxtaposition of words
of the code $\{(c^{\mu(n)}_i, D_i): i=1,\cdots,n^2\}$, defined as
in  our proof for  (\ref{t1}) above, and words
of the code
$\{(x_{j,l}^{(i),(t^n)},  D_{x_{j,l}^{(i),(t^n)}}^{Q^n}):j=1,\cdots,J_n,
l=1,\cdots,L_{t^n}\}$,  with following property

\begin{equation}P_e(C^{det}_{t^n}, t^n) < \epsilon\text{ .}\label{q1'}\end{equation}

 and for all $i\in \{1,\cdots,n^2\}$ and all $j' \in \{1,\cdots,J_n\}$:

\begin{equation}\left\| \frac{1}{L_{t^n}} \sum_{l=1}^{L_{t^n}} \sigma_{x_{j',l}^{(i),(t^n)},t^n} -
  \frac{1}{J_n} \sum_{j=1}^{J_n}  \frac{1}{L_{t^n}}\sum_{l=1}^{L_{t^n}} \sigma_{x_{j,l}^{(i),(t^n)},t^n}
  \right\|_1 <3\eta\text{ .}\end{equation}

By Lemma \ref{fann}, we have for all $t^n \in \Theta^n$ and all $i\in \{1,\cdots,n^2\}$ \[
\chi\left(W; \left\{ \left[ \frac{1}{L_{t^n}}\sum_{l=1}^{L_{t^n}}
 \sigma_{x_{j,l}^{(i),(t^n)},t^n}\right]: j=1, \cdots ,J_n\right\} \right) \leq 3\eta \log d -3\eta \log 3\eta\text{ .}\]

Therefore for any $\zeta>0$ we can choose such $\eta$ that for all $i\in \{1,\cdots,n^2\}$,
for all $t^n \in \Theta^n$
\begin{equation}\chi\left(W, \left\{ \left[  \frac{1}{L_{t^n}}\sum_{l=1}^{L_{t^n}}
\sigma_{x_{j,l}^{(i),(t^n)},t^n}\right]: j=1, \cdots ,J_n\right\} \right) <
\zeta\label{q2'}\text{ .}
\end{equation}

By (\ref{q1'}) and (\ref{q2'}), 
we see that for any distribution $P$
on $\mathcal{X}$ and any positive $\delta$, we can find a $(n,\epsilon)$-code with
secrecy rate 
$\min_{Q\in\mathcal{Q},t^n\in\Theta^n}\left[
\chi(P,\{\rho^Q_x: x\in\mathcal{X}\}) -
\frac{1}{n}\chi(P^n,\{\sigma_{x^n,t^n}: x^n\in\mathcal{X}^n\})\right]-\delta$. Therefore (\ref{t2}) holds.\vspace{0.3cm}

Is $\{\rho_{x,t}, x\in\mathcal{X}\}$ symmetrizable, then by
\cite{Ahl/Bli}, even in the case without wiretapper (we have only
the arbitrarily varying channel $\{\rho_{x,t}:
x\in\mathcal{X},t\in\Theta\}$ instead of the pairs $\{(\rho_{x,t},
\sigma_{x,t}): x\in\mathcal{X},t\in\Theta\}$), the capacity is equal
to $0$. Since  we cannot exceed the secrecy capacity of the worst
wiretap channel, (\ref{t3}) holds.
\end{proof}

\end{document}